# Sign-tunable anisotropic magnetoresistance and electrically detectable dual magnetic phases in a helical antiferromagnet

Jong Hyuk Kim,[*] Hyun Jun Shin,[*] Mi Kyung Kim,[*] Jae Min Hong, Ki Won Jeong, Jin Seok Kim, Kyungsun Moon, Nara Lee, and Young Jai Choi

**Abstract**

The helimagnetic order describes a non-collinear spin texture of antiferromagnets, arising from competing exchange interactions. Although collinear antiferromagnets are elemental building blocks of antiferromagnetic (AFM) spintronics, the potential of implementing spintronic functionality in non-collinear antiferromagnets has not been clarified thus far. Here, we propose an AFM helimagnet of $EuCo_2As_2$ as a novel single-phase spintronic material that exhibits a remarkable sign reversal of anisotropic magnetoresistance (AMR). The contrast in the AMR arises from two electrically distinctive magnetic phases with spin reorientation driven by magnetic field lying on the easy-plane, which switches the sign of the AMR from positive to negative. Further, various AFM memory states associated with the evolution of the spin structure under magnetic fields were identified theoretically, based on an easy-plane anisotropic spin model. The results reveal that non-collinear antiferromagnets hold potential for developing spintronic devices.

[*]These authors contributed equally to this work.
Correspondence: Nara Lee (eland@yonsei.ac.kr) or Young Jai Choi (phylove@yonsei.ac.kr)
Department of Physics, Yonsei University, Seoul 03722, Korea

**Introduction**

Advances in antiferromagnetic (AFM) spintronics have revealed new avenues for spin-based devices[1-3], and collinear antiferromagnets are generally utilized as the elementary basis for spintronic functionalities[4-6]. However, recent examinations of intriguing and unexpected physical phenomena in non-collinear antiferromagnets have expanded the scope of target materials for exploiting the potential of AFM spintronics[2,7-9]. In particular, a large anomalous Hall effect was observed in such materials, despite the vanishingly small magnitude of magnetization[7,10]. The Hall effect originates from the Berry curvature associated with topologically non-trivial spin textures[11-13]. Although difficult to observe experimentally, recent magnetic imaging techniques such as single spin relaxometry and scanning thermal gradient microscopy can be used to image non-collinear AFM textures and domain structures[14-16].

The ability to control and detect AFM memory states is imperative for AFM spintronics, and magnetocrystalline anisotropy has been exploited as generic foundation to manipulate AFM states[17,18]. Controlled anisotropy implies an exceptional opportunity for extensive spintronic applications[17,19]. Therefore, anisotropic magnetoresistance (AMR) has been adopted to detect various resistive states associated with crystal axes[20-23]. However, in many cases, an intricate stacking geometry with additional reference layers is required for unified spintronic functionality[24,25].

Helimagnets have a prototypical non-collinear spin structure in which the spin direction is rotated spatially in the plane, but the rotation axis is parallel to the propagation direction[26]. Helimagnets are non-collinear antiferromagnets that correspond to the zero net moment inherent in rotating spins. Thus, helimagnets offer the same advantages as antiferromagnets, such as the absence of a stray field and the features of intrinsically fast spin dynamics[27-29]. Despite these merits, a crucial issue for non-collinear AFM spintronics is the establishment of controllable factors and mechanisms for the anisotropy of non-collinear spin configurations. Specifically, the lack of a comprehensive understanding of AMR affects its manipulation and application in spintronic devices. In this work, we show that the helix-to-fan transition induces a sign-reversal of the AMR in a helical antiferromagnet of $EuCo_2As_2$ (ECA). The electrically discernible dual magnetic phases that give rise to changeable sign of the AFM memory state were verified both experimentally and theoretically. In addition, we can identify diverse AFM memory states relevant to the development of the spin structure under magnetic fields, which facilitates AFM spintronics based on non-collinear antiferromagnets.

## Materials and methods

### Sample preparation

ECA single crystals were grown by the flux method with Sn flux[30]. Eu (99.9%, Alfa Aesar), Co (99.5%, Alfa Aesar), As (99.999%, Sigma Aldrich), and Sn (99.995%, Alfa Aesar) were mixed at a 1.05:2:2:15 molar ratio of Eu:Co:As:Sn. The mixture was placed in an alumina crucible sealed in an air-evacuated quartz tube. In a high-temperature furnace, the quartz tube was dwelled at 1050 °C for 20 h, slowly cooled to 600 °C at a rate of 3.75 °C/h, and then cooled to room temperature at a rate of 100 °C/h to obtain crystals with typical dimensions of $1.5 \times 1.5 \times 0.1$ mm$^3$.

### Scanning transmission electron microscopy (STEM) measurement

We prepared ECA samples with a cutting plane perpendicular to the *a*-axis, utilizing a dual-beam focused ion beam system (Helios 650, FEI). Along the cutting plane, the images presented a well-discernible atomic structure. To minimize damage to the sample, the acceleration voltage conditions were reduced gradually from 30 to 2 keV. Dark-field images were obtained using the STEM (JEM-ARM200F, JEOL Ltd, Japan) at 200 keV with a Cs-corrector (CESCOR, CEOS GmbH, Germany) and a cold field emission gun. The size of the electron probe was 83 pm, and the range of the high-angle annular dark-field detector angle was varied from 90 to 370 mrad.

### Magnetic and transport property measurements

The temperature and magnetic field dependences of the magnetization measurements were conducted with magnetic fields along the *a*- and *c*-axes using a vibrating sample magnetometer module in a physical property measurement system (PPMS, Quantum Design, Inc.). Electric transport measurements were performed using the conventional four-probe method in the PPMS. The AMR was measured by rotating the magnetic field in the *ac* plane in a PPMS equipped with a single-axis rotator.

### Theoretical calculations

The easy-plane anisotropic spin model can be expressed as

$$\mathcal{H}/N = J_1 \sum_{i=1}^{5} \vec{S}_i \cdot \vec{S}_{i+1} + J_2 \sum_{i=1}^{5} \vec{S}_i \cdot \vec{S}_{i+2} - g\mu_B \vec{H} \cdot \sum_{i=1}^{5} \vec{S}_i + K_\theta \sum_{i=1}^{5} \cos^2\theta_i - K_5 S (\sin\theta)^n \sum_{i=1}^{5} \cos 5\varphi_i,$$

where $N$ denotes the number of Eu$^{2+}$ moments in a single layer; $J_1$ and $J_2$ represent the AFM

coupling strength between Eu$^{2+}$ moments of the two nearest layers and the next-nearest layers, respectively; $S = 7/2$ for Eu$^{2+}$ ions; $g = 2$; and $K_\theta$ denotes the magnetocrystalline anisotropy constant[31]. We consider helical spin structure of the ECA as commensurate ($k = 0.8$) for convenience of calculation and thus include only five layers with periodic boundary condition. The first and second terms indicate competing exchange interactions for an AFM helical state. The third term designates the Zeeman energy, where the magnetic field $\vec{H}$ lies on the *ac* plane, forming an angle $\theta$ with the *c*-axis. The fourth term denotes the easy-plane magnetocrystalline anisotropy energy, which favors the planar spin orientation. The fifth term reflects that we are dealing with a three-dimensional system by considering ferromagnetic interactions within a given layer as a mean field term. For the helical AFM state, in which $\varphi = \frac{4}{5}\pi$, i.e., $J_2 = 0.31 J_1$, we estimated the following parameters by fitting the theoretical results of anisotropic magnetization to the experimental data: $g\mu_B H_m / J_1 S = 1.18$, $K_\theta = 0.35 J_1 S^2$, and $K_5 = 0.022 J_1 S$, where $H_m$ indicates the occurrence of helix-to-fan transition.

**Results**

**Structure and properties of helical ECA antiferromagnet**

A two-dimensional (2D) layered ECA helimagnet forms a body-centered tetragonal structure (I4/mmm space group)[30]. The crystal exhibits a strong 2D nature, which allows it to be mechanically exfoliated. It comprises two Co$_2$As$_2$ layers placed on opposite sides in a unit cell, separated by a magnetic Eu layer, as shown in Fig. 1a[32]. The magnetic moments of the Eu$^{2+}$ ions ($S = 7/2$ and $L = 0$) are ordered helically, whereas those of the Co ions are paramagnetic and not relevant to the magnetic ordering[30,33]. The direction of the net moment in the Eu layer rotates in the *ab* plane, propagating along the *c*-axis (Fig. 1a). The pitch of the helix is, in general, incommensurate with the lattice parameter, and thus, no two layers have the same directions of the net moments[34]. In the ECA, a slightly incommensurate helimagnetic order with a propagation vector **k** = (0, 0, 0.79) has been observed via neutron diffraction[32]. This indicates that the AFM interaction between adjacent layers is frustrated by another AFM interaction between the second-neighbor layers [31]. The STEM was used to visualize structural units of alternately arranged Eu and Co$_2$As$_2$ layers (Fig. 1b). The lattice constants were obtained using fast Fourier transformation from the STEM data with $a = 0.402$ nm and $c = 1.150$ nm, similar to those in the previous results[30,32].

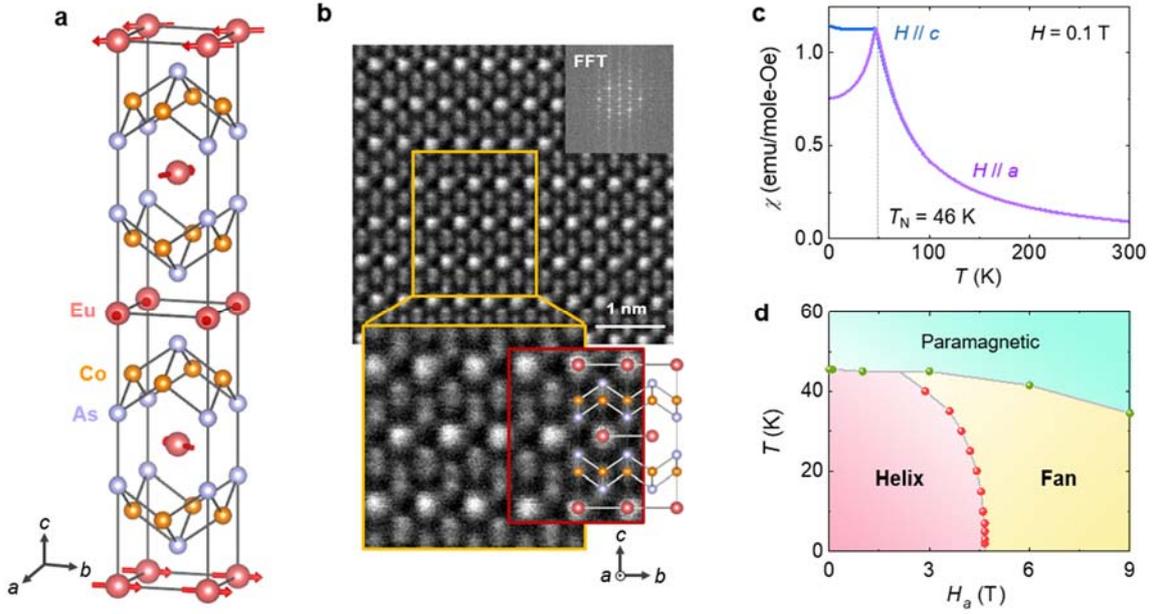

**Fig. 1 Structure and magnetic susceptibility of ECA. a** Crystal and spin structures of ECA. The red, orange, and light purple spheres denote Eu, Co, and As atoms, respectively. The red arrow on each Eu atom indicates the individual spin direction. Competing exchange interactions form the helical AFM order along the *c*-axis with ***k*** = (0, 0, 0.79). **b** Dark-field STEM images in the *bc* plane. Inset shows the diffraction pattern obtained from FFT. **c** Temperature (*T*) dependence of magnetic susceptibility $\chi$ = *M*/*H*, measured upon warming at *H* = 0.1 T after zero-field cooling for the *a*- and *c*-axes. Vertical gray line specifies the Néel *T*, $T_N$ = 46 K. **d** *H*-*T* phase diagram for *H*//*a*. Red dots distinguish the phase boundary between helix and fan phases, estimated by isothermal magnetization. The green dot indicates the Néel *T* at each *H*.

The helical AFM order emerges at $T_N$ = 46 K, as is evident from the temperature (*T*) dependence of the magnetic susceptibility defined by magnetization (*M*) divided by magnetic field (*H*), $\chi$ = *M*/*H*, measured at *H* = 0.1 T (Fig. 1c). The $\chi$ curves below $T_N$ for *H* along the *a*- and *c*-axes ($H_a$ and $H_c$) indicate an anisotropic nature. The rapid decrease of $\chi$ for $H_a$ below $T_N$ agrees with the alignment of the magnetic moments of Eu ions along the *ab* plane, as verified previously via neutron diffraction and nuclear magnetic resonance experiments[32,33]. We observed a metallic behavior with a distinct anomaly at $T_N$, as seen in the variation of the resistivity with *T* (see Supplementary Fig. S1). Therefore an indirect Ruderman–Kittel–Kasuya–Yosida (RKKY) exchange interaction between $Eu^{2+}$ spins mediated by the spins of conduction electrons can be expected[35]. Figure 1d displays a phase diagram of the *T* and $H_a$ dependences of the magnetic properties, which clarifies the phase boundary between the helix and fan structures.

**Electrically distinguishable dual magnetic phases and magnetoresistance anisotropy**

In an antiferromagnet, an enough strength of $H$ along a magnetic easy axis often generates a magnetic phase transition through spin reorientation such as a spin-flop or spin-flip transition[30,36]. The phase conversion occurs with marked anomalies in physical properties and stabilizes the flopped or flipped state. We observed a similar magnetic phase transition in the helimagnetic ECA. At 2 K, $M_a$ ($M$ along the $a$-axis) exhibits an abrupt increase at $H_m = 4.7$ T, indicative of a helix-to-fan transition[30]. The similarity of this transition to spin-flops is signified by the extrapolation of the linear slope above $H_m$ merging at the origin (Fig. 2a). $H_m$ was determined from the peak in the derivative of $M_a$. A slight magnetic hysteresis, shown in the inset of Fig. 2a, is observed, arising from the first-order characteristic of this transition (see Supplementary Information S2). Across the phase transition, a fan structure emerges in which the net magnetic moments oscillate spatially along the propagation vector[30,31,35]. By contrast, $M_c$ ($M$ along the $c$-axis) at 2 K shows a linear increase associated with the gradual canting of the net moments (Fig. 2c). The slope of $M_a$, which is smaller than that of $M_c$ below $H_m$, becomes larger across $H_m$. Therefore, the value of $M_a$ exceeds $M_c$ at $H_m$.

An easy-plane anisotropic spin model was adopted to investigate the evolution of the helix-to-fan phase. The model Hamiltonian comprises competing exchange interactions, Zeeman energy, magnetocrystalline anisotropy, and mean-field terms (see Methods and Supplementary Information S2 for details). The commensurate helical spin structure was considered for convenience of calculation. The relative angle of the two moments between the nearest layers at zero $H$ is $\varphi = \frac{4}{5}\pi$, corresponding to $J_2 = 0.31 J_1$. This is because the relative angle of two spins for a helical order is given by $\varphi = \cos^{-1}(-\frac{J_1}{4J_2})$[31,34]. Since the planar spin rotation is broken explicitly by applying $H_a$, the order parameter of the helical state can be expressed as $\sum_{i=1}^{5} \cos 5 \varphi_i$. Thus, the ferromagnetic interactions within a given layer can be treated by adding a mean field term. The estimated $M_a$ and $M_c$ values obtained from the experimental data are represented as dotted curves in Figs. 2a and 2c. The well-matched fitting in the absence of a four-fold rotational magnetocrystalline anisotropy in the $ab$ plane suggests the formation of helical spins independent of the planar crystalline axes in the ECA. Moreover, the theoretical estimation directly generates the spin configurations pertaining to the helix and fan phases, as illustrated in Figs. 2e and 2f. In the helix phase, as $H_a$ increases, the orientations of the net moments continually turn in the $H$ direction. Across the $H_m$, spin reorientation occurs in such a manner that the moments away from the $H$ direction tend to be close together perpendicularly. A further increase in $H_a$ generates an additional canting of the magnetic moments.

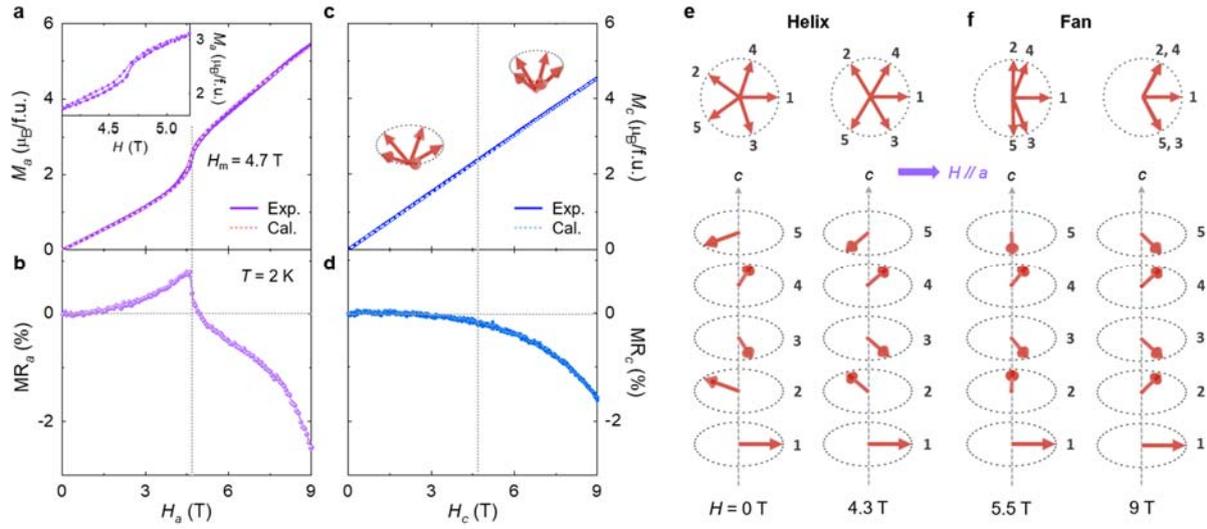

**Fig. 2 Magnetization and magnetoresistance anisotropies. a** Isothermal magnetization along the *a*-axis ($M_a$) at $T = 2$ K. Experimental and calculated data are plotted as solid and dotted curves, respectively. Vertical gray line denotes the occurrence of helix-to-fan transition. $H_m = 4.7$ T. Inset shows a magnified view of $M_a$ to elucidate magnetically hysteretic behavior. **b** Isothermal magnetoresistance MR (%) along the *a*-axis ($MR_a$) at $T = 2$ K. **c** Isothermal magnetization along the *c*-axis ($M_c$) at $T = 2$ K. Inset displays schematics of canted helical-spin structures at $H_c$, in which beginnings of the arrows in layers indicating net magnetic moments are placed together at one point. **d** Isothermal magnetoresistance MR (%) along the *c*-axis ($MR_c$) at $T = 2$ K. **e** Estimated spin configurations of helix states formed below the phase transition ($H < H_m$). **f** Estimated spin configurations of fan states formed above the transition ($H > H_m$). The red arrow indicates the net magnetic moment in each Eu layer. Each Eu layer is numbered from 1 to 5 along the *c*-axis.

The influence of anisotropic $M$ on transport was examined by magnetoresistance, MR = $\frac{R(H)-R(0)}{R(0)}$ for the *a*- and *c*-axes ($MR_a$ and $MR_c$). The increase in $M_a$ with increasing $H$ below $H_m$ (Fig. 2a) indicates a partial and gradual alignment of the magnetic moments in the $H$ direction. The average net magnetic moment along the $H$ direction disrupts the equally distributed angles between the moments in the layers (Fig. 2e). The spin configuration away from the helix state causes an increase in resistance, i.e., a positive $MR_a$ (Fig. 2b). With a further increase in $H$, $MR_a$ exhibits an abrupt reduction with the maximum slope at $H_m$ and changes to negative values. The spin-reorientation-driven switching from positive to negative $MR_a$ agrees with the crossing behavior of $M_a$. In the high-$H$ regime, $MR_a$ decreases faster with additional canting of the moments toward the flipped state (Fig. 2f). By contrast, MR$_c$ decreases as $H$ increases (Fig. 2d). This intimate correlation between the $M$ and MR plots suggests that the magnetic order governs the magnetotransport and its anisotropy. As $T$ is increased, $H_m$ decreases and the shape of the anomaly is broadened, as displayed in the anisotropic $M$ and MR plots at various $T$ values in Supplementary Fig. S4.

To probe the magnetotransport property theoretically, we propose that the interlayer hopping amplitude can be expressed by $t_{i,i+1} = t_0 + t_s |\langle \hat{n}_i | \hat{n}_{i+1} \rangle|$ ($i$ = 1-5) with periodic boundary condition, where $t_0$ and $t_s$ indicate the spin-independent and spin-dependent parts, respectively. Here, $|\langle \hat{n}_i | \hat{n}_{i+1} \rangle|$ denotes the overlap integral between the two spinors. Each spinor aligns with the direction of the net magnetic moment in each layer. The overlap integral is given by $\cos \frac{\gamma}{2}$, where $\gamma$ is the relative angle between the two spinors. For the MR calculations, the conductance ($\sigma$) of the system was assumed to be proportional to the geometric mean of multiple hopping amplitudes through the layers as $\sigma \propto \left( \prod_{i=1}^{5} t_{i,i+1} \right)^{\frac{1}{5}}$. Here the geometric mean effectively takes the average of the product $\prod_{i=1}^{5} t_{i,i+1}$. According to the definition of MR and $R = 1/\sigma$, MR is proportional to $\sigma(0) - \sigma(H)$. By setting $t_0/t_s = 0.2$, the highly anisotropic trend between MR$_a$ and MR$_c$ is moderately described by the purely spintronic consideration of the $\sigma$ (Figs. 3a and 3b). This is due to the $L = 0$ condition for the ground state of the Eu$^{2+}$ ions[34]. The MR response to an applied $H_a$ clearly distinguished two different magnetic phases. In the helix phase, the increased $\gamma$ for some of the adjacent layers at $H_a$ below $H_m$ contributes more to $\sigma$ as it decreases, thus inducing positive MR$_a$. For small $H_a$, one can explicitly prove that $\sigma$ decreases with $H_a$. In this regime, an interlayer hopping amplitude can be approximately written by $t_{i,i+1} \cong \bar{t} e^{-\alpha \epsilon_i / 2 - \beta \epsilon_i^2 / 4}$, where $\epsilon_i = \varphi_{i+1} - \varphi_i - \frac{4\pi}{5}$ represents the small deviation of the relative spin orientation $\varphi_{i+1} - \varphi_i$ from the average value, $\bar{t} = t_0 + t_s \cos \frac{2\pi}{5}$, $\alpha = t_s \sin \frac{2\pi}{5} / \bar{t}$, and $\beta = t_s \left( t_s + t_0 \cos \frac{2\pi}{5} \right) / 2 \bar{t}^2$. Since $\sum_{i=1}^{5} \epsilon_i = 0$, $\sigma \propto \left( \prod_{i=1}^{5} t_{i,i+1} \right)^{\frac{1}{5}} = \bar{t} e^{-\beta \sum_{i=1}^{5} \epsilon_i^2 / 20} < \bar{t}$. Hence $\sigma$ decreases for small $H_a$. In the fan phase, $\gamma$ decreases continuously with a further increase in $H_a$ by approaching to the completely spin aligned state along the magnetic field direction, which enhances the $\sigma$ and causes MR$_a$ to reduce. Various AFM memory states relevant to the different spin configurations under an external $H_a$ were identified theoretically, as shown in the schematics of Fig. 3a.

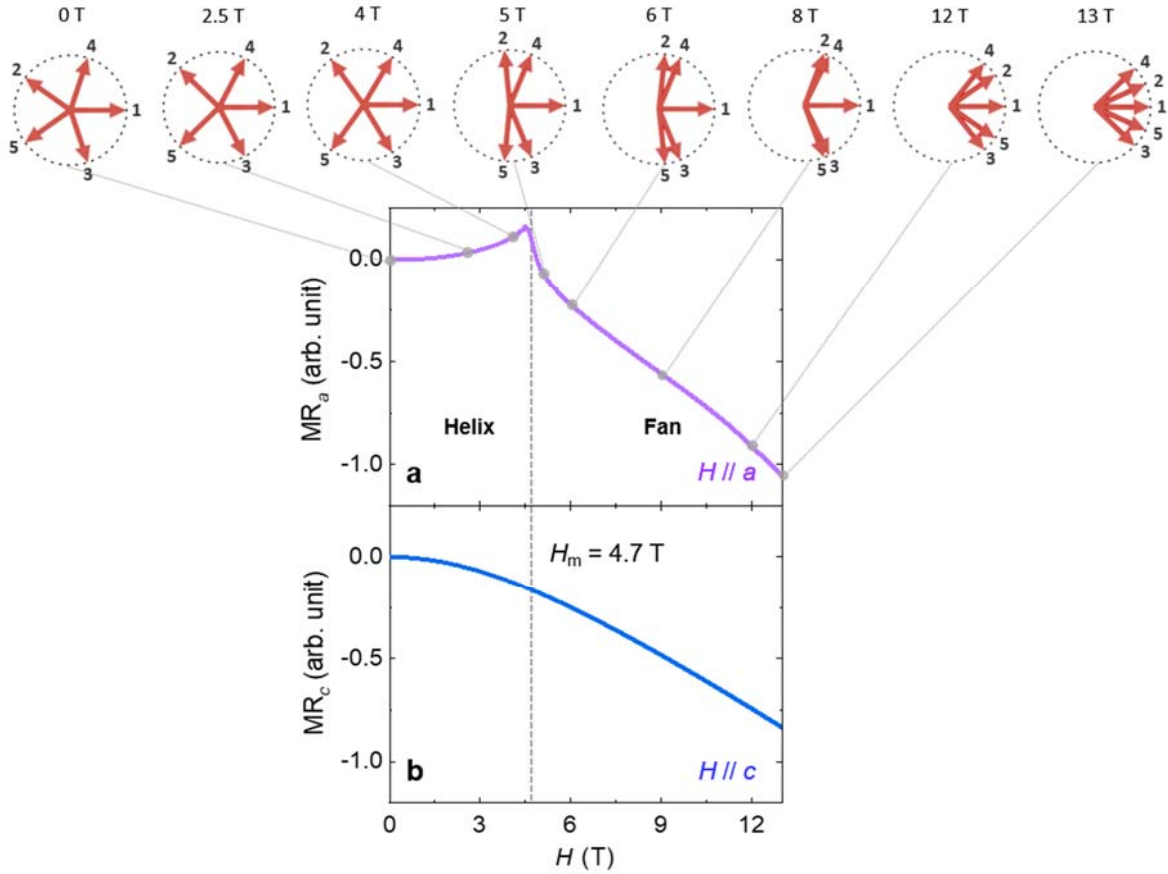

**Fig. 3 Calculated magnetoresistance anisotropy. a** Theoretically estimated isothermal magnetoresistance, MR, along the *a*-axis ($MR_a$). Schematics of net magnetic moment configurations correspond to different $MR_a$ states at various $H_a$'s. **b** Theoretically estimated isothermal magnetoresistance, MR, along the *c*-axis ($MR_c$).

**Spin-reorientation-driven reversal of anisotropic magnetoresistance**

A peculiar spintronic characteristic of non-collinear AFM ECA is presented by angle-dependent magnetotransport. The results of the AMR, defined as $\frac{R(\theta)-R(0)}{R(0)}$, are shown in Figs. 4a and 4b. As indicated in the geometry of the AMR measurement in Fig. 4(a), $H$ is continually rotated perpendicular to the current, excluding the extrinsic Lorentzian MR effect. At $H < H_m$, the *ab*-planar helix formation allows for two-fold rotational symmetry. The AMR is maximized at $\theta = 90°$ and $270°$ (Fig. 4a) due to the positive value of $MR_a$, which increases gradually with increasing $H$ (Fig. 2b). At $H$ values slightly exceeding $H_m$, the AMR near $\theta = 90°$ and $270°$ begins to reverse partially, which is observed as a dip in Fig. 4a. A further increase in $H$ leads to a complete reversal of the AMR. The AMR contour map (Fig. 4b) demonstrates the sign-tunable AMR across $H_m$. The detailed influence of increasing $T$ on the AMR is plotted in the contour maps of Supplementary Fig. S5. The overall trend of AMR development under the applied *H* was reproduced using theoretical calculations, as shown in Figs. 4c and 4d. The contrast emerging from the reversal behavior of the AMR effect reflects the intrinsic bulk properties and clarifies the different magnetotransport features between the helix and fan structures.

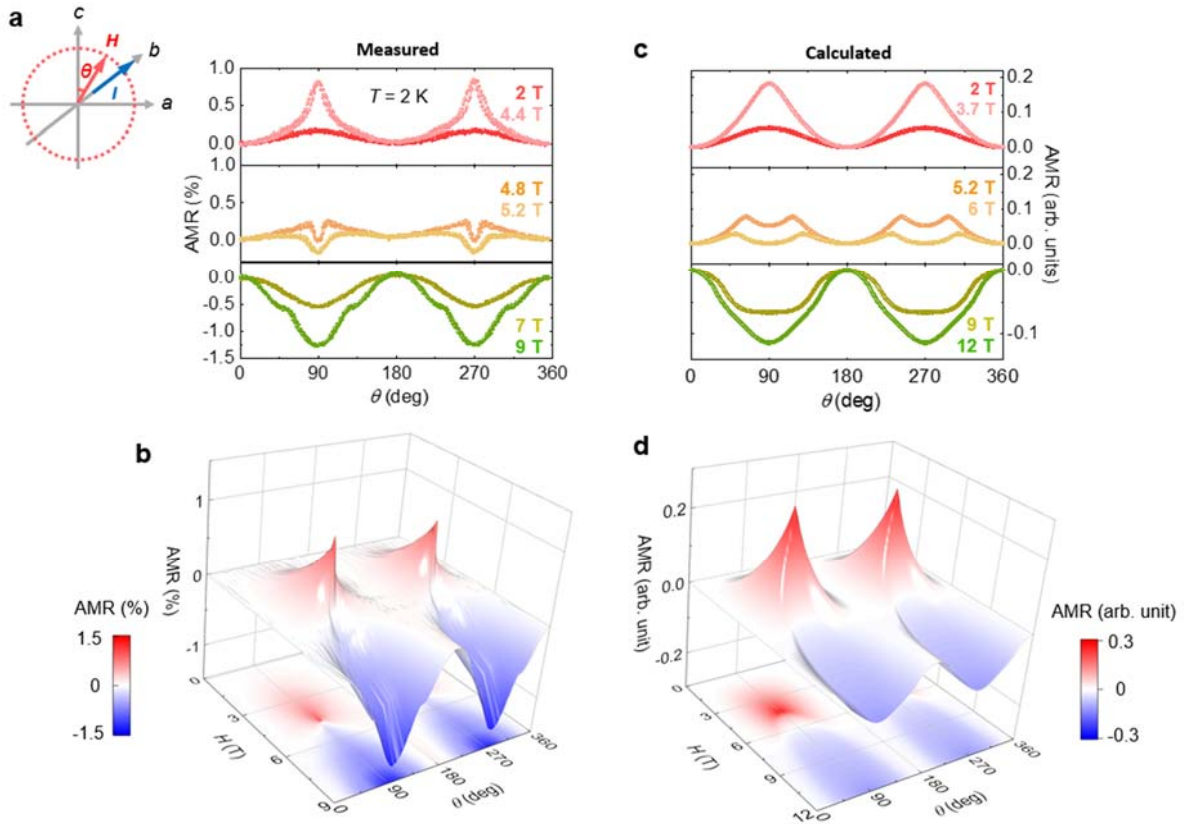

**Fig. 4 Experimental and theoretical anisotropic magnetoresistance. a** Plots of the AMR (%) measured at *T* = 2 K by rotating *H* =2, 4.4, 4.8, 5.2, 7, and 9 T in the *ac* plane with the current along the *b*-axis, *I*//*b*. Geometry of the AMR measurement is schematically shown. $\theta = 0°$ for the *c*-axis and $\theta = 90°$ for the *a*-axis. **b** Contour plots of the AMR at *T* = 2 K. **c** Plots of the calculated AMR data at *H* =2, 3.7, 5.2, 6, 9, and 12 T. **d** Contour plots of the AMR obtained from theoretical estimations.

**Discussion**

We have demonstrated a mechanism to generate the sign-changing AMR phenomenon originating from electrically distinctive dual magnetic phases separated by magnetic phase transition. The magnetic transition emerges from the competition of diverse energy scales included in the model Hamiltonian. Specifically, the highly 2D characteristic of magnetocrystalline anisotropy plays an important part in the good agreement between bulk measurements and theoretical calculations. In pursuance of the AMR effect caused predominantly by the magnetocrystalline anisotropy in AFM spintronics, the approach proposed in this study is applicable to a variety of antiferromagnets in which the feature of anisotropic magnetotransport would appear differently relying on the change of magnetocrystalline anisotropy. The development of magnetic devices with the desired properties requires detectable macroscopic effects that are contingent upon the variable magnetic states. In complex non-collinear antiferromagnets, more spintronic effects can be present. Therefore, the electrical access to various AFM memory states associated with the evolution of helical spin texture would provide an opportunity for the realization of multi-level AFM memory devices.

Crucially, the reversal phenomenon of the AMR has rarely been reported before. In a previous work on ferromagnetic $La_{0.7}Ca_{0.3}MnO_3$ ultra-thin films, a sign reversal of the AMR was shown to originate from the planar tensile strain that facilitates the rotation of the ferromagnetic easy axis, which is different from the present case[37]. The sign-tunable AMR effect in a single-phase non-collinear antiferromagnet is unique, owing to its intrinsic origin from the electrically discernible magnetic phases. Additionally, it has been known that a variety of topological states mediated by spin-orbit interaction gives rise to exotic topological magnetism[38-40]. Very recently, a helical magnetism driven by Weyl-mediated RKKY interactions was observed in a Weyl semimetal, NdAlSi[41]. The recognition of versatile AFM memory states in a helimagnet offers valuable guidelines for investigating the intimate interplay between electronic and magnetic topological properties and thus for implementing topological AFM spintronics.

In summary, we present a new single-crystalline spintronic material in which the sign-tunable AMR effect reflects fully intrinsic bulk properties. Our results are useful in the context of the development of AFM spintronics, which has been driven by novel materials. Theoretically, the highly anisotropic 2D spin characteristic was highlighted as a core factor for anisotropic magnetotransport in a natural non-collinear AFM ECA. The scheme used in this work is a particular type of spin-reorientation-driven mechanism to study intriguing anisotropic properties, which can motivate further investigations of non-collinear AFM for extensive spintronic applications.

**Acknowledgments**
This work was supported by the National Research Foundation of Korea (NRF) through grants

**Author details**

Department of Physics, Yonsei University, Seoul 03722, Korea

**Conflict of interest**

The authors declare that they have no conflict of interest.

**Supplementary information** The online version contains supplementary material

# Supplementary Information

# Sign-tunable anisotropic magnetoresistance and electrically detectable dual magnetic phases in a helical antiferromagnet

Jong Hyuk Kim,[*] Hyun Jun Shin,[*] Mi Kyung Kim,[*] Jae Min Hong, Ki Won Jeong, Jin Seok Kim, Kyungsun Moon, Nara Lee, and Young Jai Choi

*Department of Physics, Yonsei University, Seoul 03722, Korea*

*These authors contributed equally to this work.

Correspondence: Nara Lee (eland@yonsei.ac.kr) or Young Jai Choi (phylove@yonsei.ac.kr)
Department of Physics, Yonsei University, Seoul 03722, Korea

**S1. Metallic behavior of ECA**

ECA belongs to the $ThCr_2Si_2$-type structure family and forms an un-collapsed tetragonal structure[1]. Under high pressures exceeding 4.7 GPa, the formation of considerable As-As bonds along the *c*-axis leads to a structural collapse associated with a mixed valence state of Eu ions[2]. The further oxidation of Eu ions involves electron doping into the Co ($3d$) sub-band, which results in the ferromagnetic order of both Eu and Co ions at $T_C$ = 125 K[2]. The ECA single crystals exhibit a helical antiferromagnetic (AFM) order at $T_N \approx 46$ K (Fig. 1c in the main manuscript). The *T* dependence of resistivity ($\rho$) was also measured using the conventional four-probe method. The $\rho$ as a function of *T* reveals a metallic behavior as it decreases with decreasing *T* (Fig. S1a). Although a small kink was observed at $T_N$, the anomalous behavior at $T_N$ was distinctly identified in its *T* derivative (Fig. S1b).

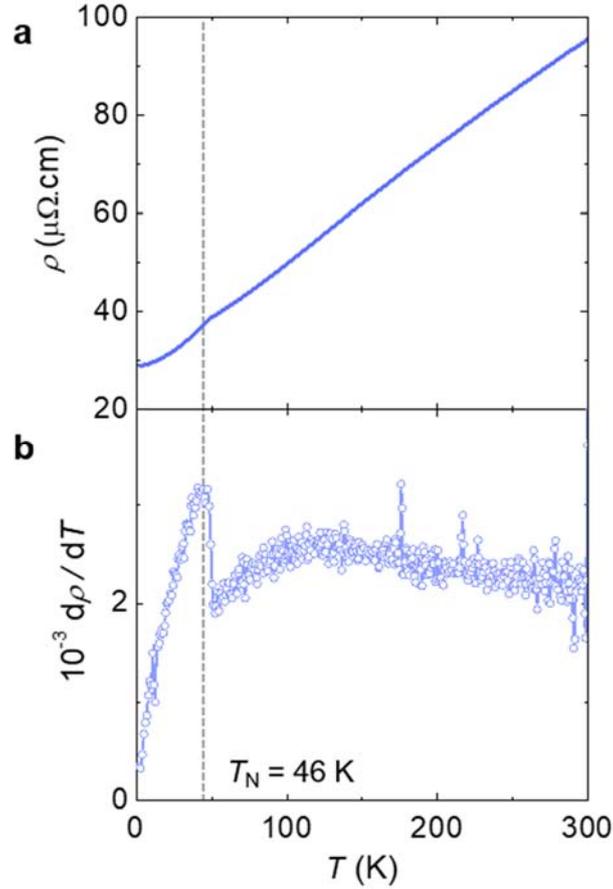

**Fig. S1 Metallic behavior of resistivity. a** Temperature dependence of resistivity, $\rho$, at zero magnetic field. Vertical grey line indicates the Néel temperature, $T_N = 46$ K. **b** Temperature derivative of $\rho$.

**S2. Easy-plane anisotropic spin model**

The ECA exhibits an AFM helical state, where the $Eu^{2+}$ moment rotates by approximately $\varphi = \frac{4}{5}\pi$ per layer, propagating along the $c$-axis[1]. When $H_a$ is applied, it is known that the planar spin system undergoes a phase transition from the AFM helical to the fan state[3], and within a layer, the spins are ferromagnetically coupled. As shown in Fig. S2, $J_1 = 4J_B$ and $J_2 = J_C$ represent the AFM couplings between $Eu^{2+}$ moments between the two nearest and next-nearest layers, respectively[3]. We consider helical spin structure of the ECA as commensurate ($k = 0.8$) for convenience of calculation and thus include only five layers with periodic boundary condition.

The tetragonal symmetry of the ECA also allows a four-fold rotational magnetocrystalline anisotropy in the *ab* plane. The formation of helical spins independent of the *ab* planar crystalline axes in the ECA implies that this term is negligible. The planar spin rotation symmetry is broken explicitly by applying $H_a$, and this allows us to express the order parameter of the helical state as $Q_5 = \sum_{i=1}^{5} \cos 5\varphi_i /5$. As expressed in the model Hamiltonian, the ferromagnetic interactions in a given layer can be treated by adding a mean field term, which couples to the order parameter $Q_5$. D. C. Johnston has assumed that the ferromagnetic couplings in a layer completely aligns the spins in the same direction[3]. The fact that we are actually dealing with a three dimensional system has been reflected in this mean-field approach. By contrast, when $H_c$ is applied, the planar spin rotation symmetry is spontaneously broken, and there exists a gapless Goldstone mode. Hence, it is not appropriate to add such a mean-field term along this direction. We have added a phenomenological factor $(\sin\theta)^n$ to account for this. We selected $n = 4$ and checked if the result was insensitive to the choice of $n > 4$. In Fig. S3a, the ground state energy $E/J_1$ of the spin Hamiltonian is plotted as a function of $H_a$. It exhibits a cusp at $H_m$, which demonstrates that the phase transition from the AFM helical to the fan state is a first-order. In Fig. S3b, the order parameter $Q_5$ is plotted as a function of $H_a$. It clearly demonstrates a finite discontinuity in the order parameter $Q_5$ at $H_m$.

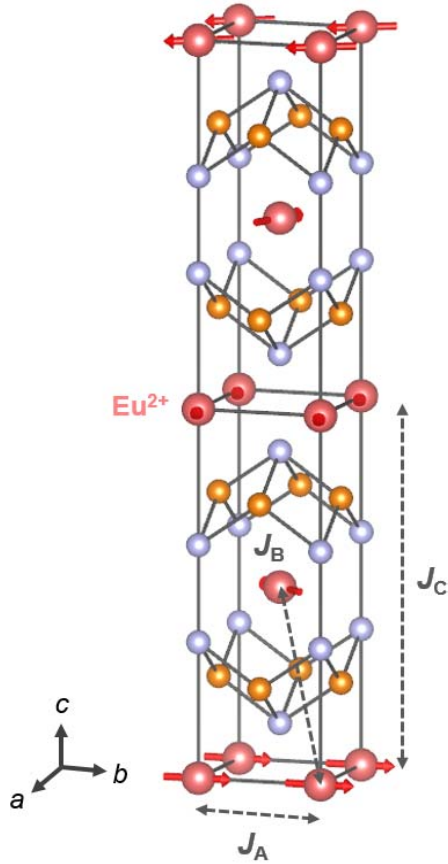

**Fig. S2 Competing exchange interactions.** Magnetic exchange interactions are defined in the structural figure. $J_A$ is the ferromagnetic coupling within an $Eu^{2+}$ layer. $J_B$ and $J_C$ are AFM couplings for nearest and next-nearest layers, respectively.

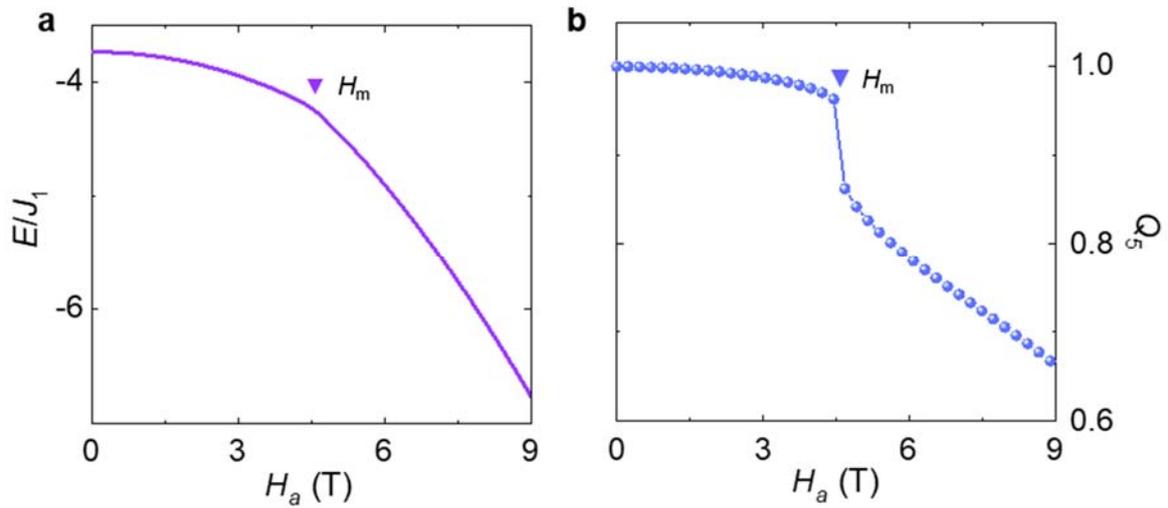

**Fig. S2 Ground state energy and order parameter.** **a** Ground state energy $E$ is plotted as a function of $H_a$ in units of $J_1$. **b** Order parameter $Q_5$ is plotted as a function of $H_a$.

**S3. Temperature evolution of isothermal magnetization and magnetoresistance**

At 2 K, spin reorientation occurs across the helix-to-fan transition at $H_m$ = 4.7 T, resulting in the phase conversion from a helical to a fan structure (Fig. 2a in the main manuscript)[3]. As $T$ increases, the noticeable anomaly of $M_a$ at 2 K is progressively reduced and broadened (Fig. S4a). $H_m$ is determined by the $H$-derivative of $M_a$, and it is lowered from 4.5 T at 10 K to 2.8 T at 40 K. The slope of $M_a$ below $H_m$ continues to increase as $T$ increases, which suggests a thermally diminished stiffness of helical spins. In addition, the value of $M_a$ at the highest value of $H$ = 9 T shows a continual decrease with increasing $T$, owing to thermal fluctuation. By contrast, the slope of the linearly increased $M_c$ due to the gradual canting of $Eu^{2+}$ spins remains unchanged up to 30 K (Fig. S4b), implying that the magnetically hard $c$-axis can be thermally sustained. At $T$ = 40 K, the linear increase in $M_c$ is maintained up to ~7 T, above which the slope of $M_c$ is reduced to some extent. A similar $T$ evolution was observed in the magnetoresistance anisotropy. The peaky feature of $MR_a$ at $H_m$ = 4.7 T and $T$ = 2 K (Fig. 2b in the main manuscript) decreases continually in conjunction with the lowered $H_m$ as $T$ increases (Fig. S4c). However, the monotonously decreasing tendency of $MR_c$ is relatively well preserved at higher $T$ (Fig. S4d).

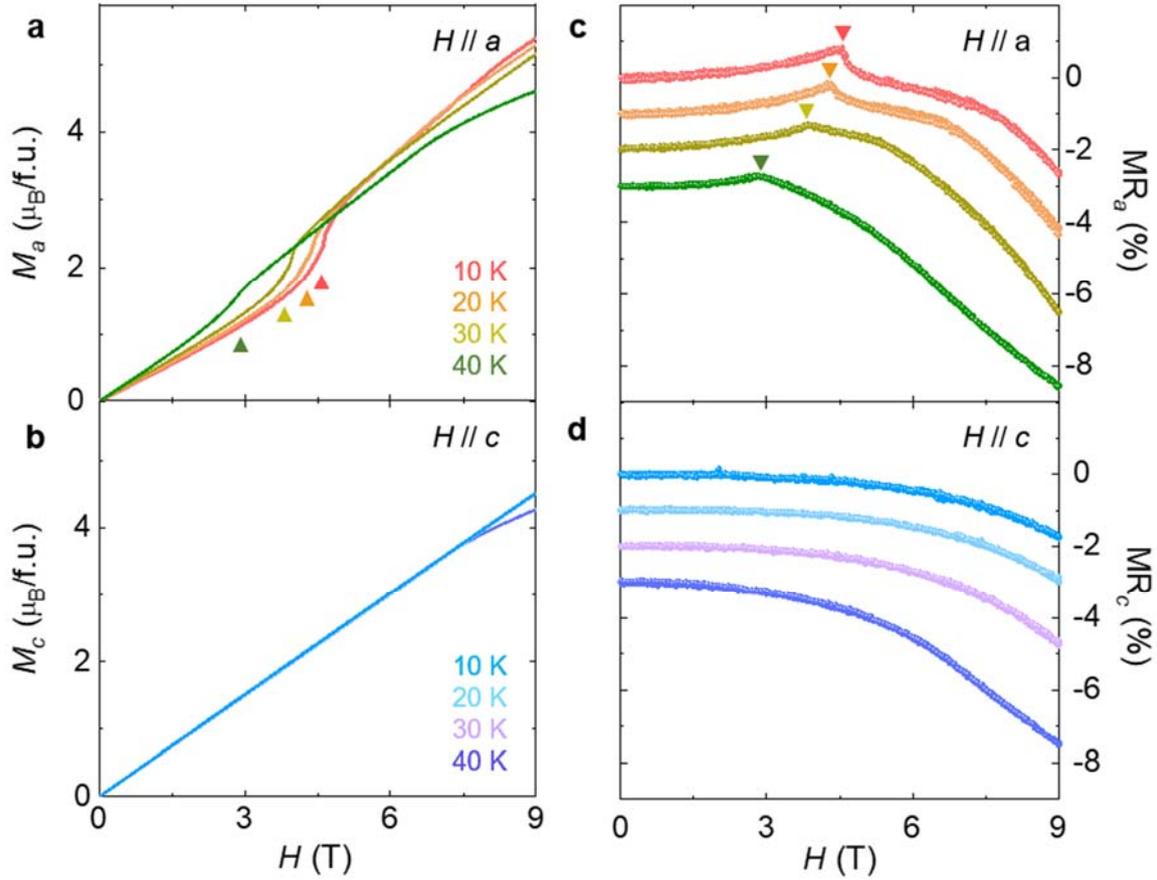

**Fig. S4 Temperature evolution of isothermal magnetization and magnetoresistance. a** Isothermal $M_a$ at $T$ = 10, 20, 30, and 40 K. Closed triangle denotes the occurrence of the helix-to-fan transition, $H_m$, at each $T$. **b** Isothermal $M_c$ at $T$ = 10, 20, 30, and 40 K. **c** $MR_a$ at $T$ = 10, 20, 30, and 40 K. The inverted triangles indicate the occurrence of $H_m$ at each $T$. **d** $MR_c$ at $T$ = 10, 20, 30, and 40 K. The MR data are shifted vertically for clear visualization.

**S4. Temperature evolution of anisotropic magnetoresistance**

We examined the $T$ evolution of the anisotropic magnetoresistance, AMR = $\frac{R(\theta)-R(0)}{R(0)}$, measured by rotating $H$ in the *ac* plane with the current along the *b*-axis. The AMR at 10 K shows a reversal behavior similar to that measured at 2 K, as shown in the contour map of Fig. S5a. The occurrence of the helix-to-fan transition is slightly lowered as $H_m$ = 4.5 T. At 20 and 30 K (Figs. S5b and S5c), the sign reversal of AMR occurs at a significantly higher $H$, because $MR_a$ is reduced very slowly above $H_m$ (Fig. S4c), which is associated with the magnetic phase transition broadened as $T$ increased. At 40 K, the smaller magnitude of the $MR_a$ anomaly at $H_m$ is noted, but a relatively faster decrease in the MR$_a$ above $H_m$ causes a lower $H$ for the AMR reversal (Fig. S5d).

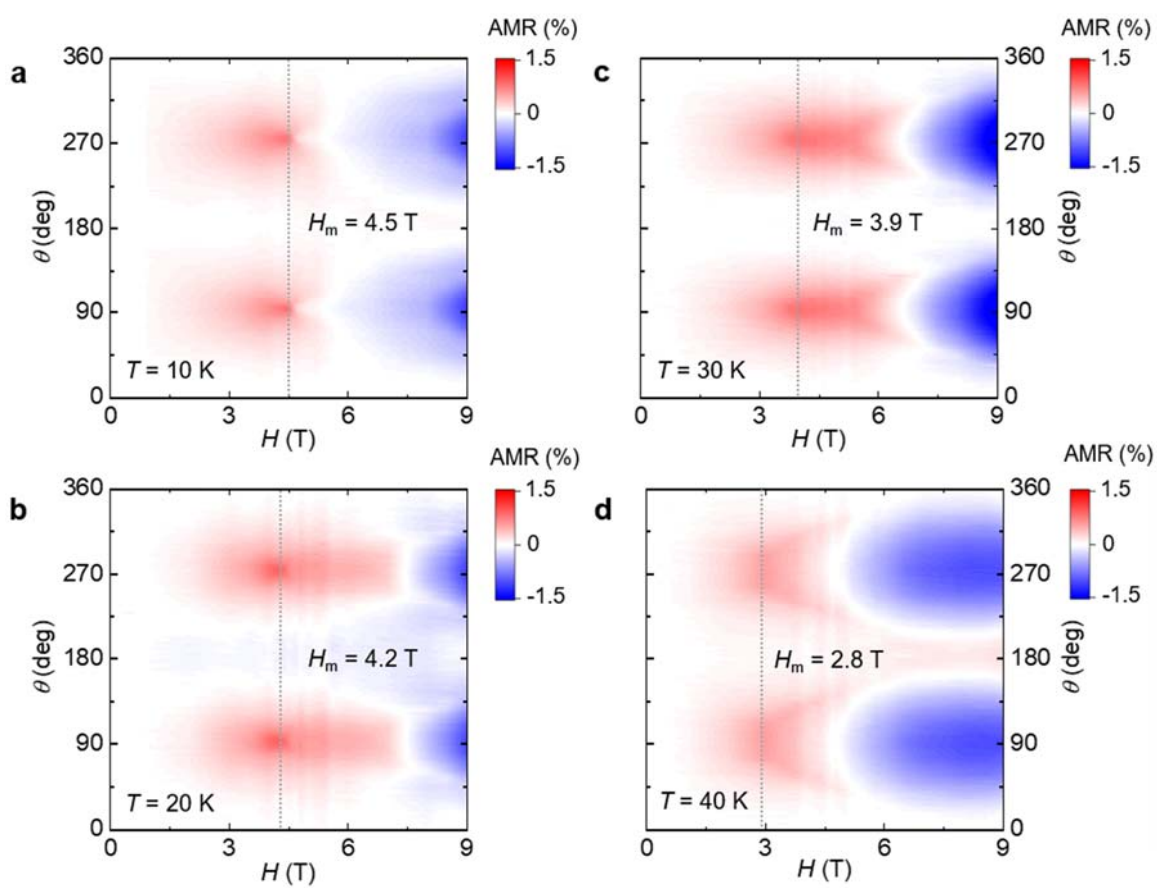

**Fig. S5 Temperature evolution of anisotropic magnetoresistance. a-d** $H$-$\theta$ contour plot established from the AMR data measured at various values of $H$ and $T$ = 10, 20, 30, and 40 K, respectively. The vertical dotted lines indicate the occurrence of helix-to-fan transition at each $T$.